\documentclass[aps,onecolumn,showpacs,preprint]{revtex4}
\usepackage{amssymb}
\usepackage{amsmath}
\usepackage{graphicx}

\newcommand{\ve}{\varepsilon}

\begin{document}

\setcounter{page}{1}

%\title[]{Edge states of bilayer zigzag graphite nanoribbons}
\title[]{Edge states of zigzag bilayer graphite nanoribbons}

%\author{Gil Dong \surname{Hong}} \email{jkps@kps.or.kr}
%\thanks{Fax: +82-2-554-1643}

\author{Jun-Won \surname{Rhim}}
\affiliation{Department of Physics and IPAP, Yonsei University,
Seoul, 120-749, Korea }
\author{Kyungsun \surname{Moon}}
\affiliation{Department of Physics and IPAP, Yonsei University,
Seoul, 120-749, Korea }
%\date{\today}

\begin{abstract}
Electronic structures of the zigzag bilayer graphite
nanoribbons(Z-BGNR) with various ribbon width $N$ are studied within
the tight binding approximation. Neglecting the inter-layer hopping
amplitude $\gamma_4$, which is an order of magnitude smaller than
the other inter-layer hopping parameters $\gamma_1$ and $\gamma_3$,
there exist two fixed Fermi points $\pm k^*$ independent of the
ribbon width with the peculiar energy dispersion near $k^*$ as $\ve
(k) \sim \pm (k-k^*)^N$. By investigating the edge states of the
Z-BGNR, we notice that the trigonal warping of the bilayer graphene
sheets are reflected on in the edge state structure. With the
inclusion of $\gamma_4$, the above two Fermi points are not fixed,
but drift toward the vicinity of the Dirac point with the increase
of the width $N$ as shown by the finite scaling method and the
peculiar dispersions change to the parabolic ones. The edge
magnetism of the Z-BGNR is also examined by solving the half-filled
Hubbard Hamiltonian for the ribbon using the Hartree-Fock
approximation. We have shown that within the same side of the edges,
the edge spins are aligned ferromagnetically for the experimentally
relevant set of parameters.
\end{abstract}



\maketitle

\section{INTRODUCTION}
Graphite nanoribbons(GNR) are the quasi one-dimensional carbon
sheets, whose width and the chirality can be well controlled by the
current nano-lithography technique.\cite{R1} Due to the high
mobility of the sample and the relativistic energy dispersion
relation of the massless Dirac fermion, there has been a lot of
interest both theoretically and experimentally to investigate the
electronic properties of the
GNR.\cite{R2,R3,R4,R5,R6,R7,R8,R9,R10,R11,R12,R13,R14,R15,R16,R17}
The electronic properties of the GNR are closely controlled by the
chirality as is well known for the case of carbon nanotube. Among
the GNRs with different chiralities, the armchair and zigzag GNR
have been studied most intensively for the
decades.\cite{R2,R3,R4,R5,R6,R9,R10,R11,R12,R13,R14,R15,R16,R17}
Unlikely from the carbon nanotube with the compact angle variable
for the transverse modes, the GNR can support the edge states.
Recently, the edge state of zigzag GNR has been of great interest
due to its peculiar dispersion relation with the almost flat edge
bands near the zero energy.\cite{R3,R14,R15} When the Coulomb
interactions are taken into account, the existence of the flat edge
bands may lead to the edge magnetism for various kinds of ribbon
edges.\cite{R3,R17,R18,R19} They have shown that the edge magnetism
of zigzag GNR is ferromagnetic along each edge while
anti-ferromagnetic between the two different edges due to the
bipartite nature of the lattice structure. This kind of edge
magnetism has been widely studied within first-principle
calculations.\cite{R20,R21,R22,R24} Furthermore, its potential as a
half-metallic material under lateral electric field has been
demonstrated within a first principle approach by Son {\em et.
 al.}.\cite{R18}
Hence to identify the edge states of the various GNRs is crucial to
understand both the magnetic and electric properties of the system.

In the paper, we study the electronic and magnetic structure of
the zigzag bilayer graphite nanoribbons with various ribbon width
$N$. First, we give a brief review of the bilayer graphene system.
The bilayer graphene system is two coupled graphene sheets and its
tight binding electronic structures are determined by four leading
characteristic hopping parameters as shown in Fig. 1(a). It has
demonstrated quite unique electronic properties compared to the
single graphene such as the larger Hall step and different kinds
of chiral quasiparticles and the distinct electronic
structures.\cite{R25} The most distinct electronic feature of
bilayer graphene system is the very existence of the trigonal
warping of the Fermi level in the vicinity of the Dirac points as
demonstrated in the inset of Fig. 2(a). It means that the band of
the graphite bilayer near the Dirac point does not form a cone but
a trigonally curved cone and there exist three additional pockets
along the directions of zigzag chirality. It has been shown that
these peculiar electronic structures, which are controlled by the
inter-layer hopping parameters $\gamma_1$ and $\gamma_3$, produce
extra degeneracies in the Landau levels. The three extra dirac
cones are known to play an important role in determining the
minimal conductivity of the graphite bilayer.\cite{R26}

In the work, we focus on the edge state structure of the Z-BGNRs
with various ribbon width $N$. Since the three legs in the band
structure of the graphite bilayer lie along the direction of zigzag
chirality, we have chosen the Z-BGNR in order to see the possible
signature of the trigonal warping imprinted on the edge state
structure of the system. The schematic structure of the Z-BGNR is
shown in Fig. 1(a). We have chosen the Bernal stacked graphite
nanoribbons, where the widths of the two layers are the same and the
width is defined to be the number of dimer lines $N$ as shown in
Fig. 1(b). We use the tight binding method to calculate the band
structures of the Z-BGNRs with varying widths. According to the
experimental data, the inter-layer hopping amplitude $\gamma_4$ is
much smaller than $\gamma_3$ by an order of magnitude.\cite{R27}
Neglecting $\gamma_4$, we notice that the Z-BGNR has two fixed Fermi
points at $k=\pm k^*$ with
$k^*=2\cos^{-1}(\sqrt{\bar{\gamma_1}\bar{\gamma_3}}/2)$ and the
following energy dispersion relation $\ve\sim\pm(k-k^*)^N$. As $N$
goes to infinity, the dispersionless edge band appears. By analyzing
the semi-infinite Z-BGNR case analytically, we have demonstrated
that the trigonal warping manifests itself in the edge band
structure of the Z-BGNR. Interestingly, the existence of the edge
states seems to be consistent with the condition imposed by the
Berry phase argument.\cite{R28} With the inclusion of $\gamma_4$,
the particle-hole symmetry is broken and the energy gap appears in
the dispersion. In this case, the conduction band minimum and the
valence band maximum occurs at two different points $k^e$ and $k^h$
respectively. They are no more fixed points independent of $N$
rather drift toward Dirac point with increasing $N$. By examining a
semi-infinite Z-BGNR with $\gamma_4$ taken perturbatively and using
the scaling method as well, we have analyzed the edge states in
further detail. We notice that one edge band remains to be flat
while the other edge band becomes dispersive. These two edge bands
appear below $\ve=0$ and split proportional to the hopping parameter
$\gamma_4$. Finally, the magnetic properties of the Z-BGNR are
studied by investigating the Hubbard Hamiltonian obtained from the
inclusion of the on-site Coulomb repulsion $U$ based on the
Hartree-Fock approximation. We have calculated the magnetic moments
at the edge as a function of $U$ for half-filled system at zero
temperature. Our main concern is that the possible half metallicity
would persist in the Z-BGNR upon lateral gate bias. Our result has
shown that for the realistic value of $U$ ($U/\gamma_0 \cong 1$),
the ferromagnetic alignment of the edge spins within the same side
of the edges is favorable. This implies that the spintronic
application of the Z-BGNR is quite promising.

The paper is organized as follows. In Sec.II, a detailed
descriptions about the tight binding model for the Z-BGNRs are
given. In Sec.III, we investigate the electronic structures of
Z-BGNRs with or without the small inter-layer hopping parameter
$\gamma_4$. We focus on the edge state structure based on the
formulae of Sec.II. Without $\gamma_4$, the coupled equation to
obtain the energy dispersion relation can be simplified and one can
extract the important features of the edge states in the system. We
analyze the semi-infinite Z-BGNRs to obtain the analytic formulae of
the edge states amplitudes. In Sec.IV, the interlayer edge magnetism
of the Z-BGNR is studied by including the Hubbard type
onsite-interactions. Finally, summaries will be given in Sec.V.

\section{TIGHT BINDING HAMILTONIAN}
We introduce the tight binding model for the Z-BGNR by considering
the four hopping parameters $\gamma_0,\gamma_1,\gamma_3$ and
$\gamma_4$, where $\gamma_0$ is the intra-layer hopping and the
others the inter-layer hopping amplitudes as shown in Fig. 1(a).
This type of approach has been successfully employed to study the
electronic structure due to the $p_z$-orbital of various carbon
based materials including the GNR.\cite{R3,R15,R16} We begin with
the following Hamiltonian
\begin{eqnarray}
H=-\sum_{\langle i,j\rangle}t_{ij}c^\dag_i c_j \label{tbinding}
\end{eqnarray}
with
\begin{eqnarray}
t_{ij}=\langle\varphi (\vec{r}-\vec{R}_i)|H|\varphi
(\vec{r}-\vec{R}_j)\rangle
\end{eqnarray}
where the summations over $i$ and $j$ include all the hopping
processes related to the four hopping parameters. For the sake of
simplicity, we redefine the index $i$ in terms of the following
three indices such as $c_i = c_{\alpha}(m,\lambda)$, where $\alpha$
represents the unit cell of Z-BGNR, $m$ the index of dimer lines
which runs from 1 to N, and $\lambda$ one of the four sublattices
$A,B,\tilde{A}$ and $\tilde{B}$ as shown in Fig. 1(b). By
transforming this to the collective modes
\begin{eqnarray}
c_{m,\lambda}(k)=\frac{1}{\sqrt{N}}\sum_\alpha e^{-ik
y_\alpha}c_\alpha(m,\lambda)
\end{eqnarray}
where $y_\alpha$ denotes the position of $\alpha$-th unit cell. The
general states can be written by
\begin{eqnarray}
|\psi
(k)\rangle=\sum_{m,\lambda}\psi_{m\lambda}c^{\dag}_{m,\lambda}(k)|0\rangle.
\label{gstate}
\end{eqnarray}
By substituting Eq. (\ref{gstate}) into Eq. (\ref{tbinding}), one
can obtain the following four recurrence relations
\begin{widetext}
\begin{eqnarray}
\varepsilon \psi_{mA}&=&\gamma_0
(2\cos\frac{k}{2}\psi_{mB}+\psi_{m-1B})+\gamma_3
(2\cos\frac{k}{2}\psi_{m-1\tilde{B}}+\psi_{m\tilde{B}})+\gamma_4
(2\cos\frac{k}{2}\psi_{m\tilde{A}}+\psi_{m-1\tilde{A}})\\
\nonumber
%\end{eqnarray}
%\begin{eqnarray}
\varepsilon \psi_{mB}&=&\gamma_0
(2\cos\frac{k}{2}\psi_{mA}+\psi_{m+1A})+\gamma_1
\psi_{m\tilde{A}}+\gamma_4
(2\cos\frac{k}{2}\psi_{m\tilde{B}}+\psi_{m-1\tilde{B}})\\
\nonumber
%\end{eqnarray}
%\begin{eqnarray}
\varepsilon \psi_{m\tilde{A}}&=&\gamma_0
(2\cos\frac{k}{2}\psi_{m\tilde{B}}+\psi_{m-1\tilde{B}})+\gamma_1
\psi_{mB}+\gamma_4 (2\cos\frac{k}{2}\psi_{mA}+\psi_{m+1A})\\
\nonumber
%\end{eqnarray}
%\begin{eqnarray}
\varepsilon \psi_{m\tilde{B}}&=&\gamma_0
(2\cos\frac{k}{2}\psi_{m\tilde{A}}+\psi_{m+1\tilde{A}})+\gamma_3
(2\cos\frac{k}{2}\psi_{m+1A}+\psi_{mA})+\gamma_4
(2\cos\frac{k}{2}\psi_{mB}+\psi_{m+1B}). \nonumber
\label{recurrence}
\end{eqnarray}
\end{widetext}

Here we use the dimensionless wave vector $k$ in which the primitive
lattice vector of the ribbon is embedded so that the range of $k$
runs from $-\pi$ to $\pi$. Since the Z-BGNR is terminated along the
$x$-axis, the following boundary conditions are imposed:
$\psi_{0\lambda}=\psi_{N+1\lambda}=0$. From the recurrence
relations, we notice that, when $\gamma_4$ is taken to be zero,
particle-hole symmetry exists in the band structures. In other
words, for $\gamma_4=0$, one set of eigenvalue and eigenvector
solution of \{$\varepsilon,
\psi_{mA},\psi_{mB},\psi_{m\tilde{A}},\psi_{m\tilde{B}}$\}
guarantees the existence of the other orthogonal solution
\{$-\varepsilon,
-\psi_{mA},\psi_{mB},-\psi_{m\tilde{A}},\psi_{m\tilde{B}}$\}. Hence
the band structure is symmetric with respect to $\ve=0$. On the
other hand, when $\gamma_4$ is included, the particle-hole symmetry
is broken.

\section{Electronic structure of the Z-BGNR}

We study the electronic structures of the Z-BGNRs with finite
ribbon width $N$ with special focus on the edge states. We first
investigate the semi-infinite Z-BGNR and analytically show the
signatures of the trigonal warping of the Fermi surface in the
edge states. Hence our main interest lies in the bands near
$\varepsilon=0$, which contain the states localized at the ribbon
edge.
%We consider the semi-infinite ribbon, that is,
%$N\rightarrow\infty$.
For $\varepsilon=0$, the four coupled equations in Eq. (5) are
decomposed into the two coupled ones. One of the two coupled
equations only involves $\psi_B$ and $\psi_{\tilde{B}}$. By solving
these equations, we notice that they lead to unphysical solutions
unless $\psi_{mB}=\psi_{m\tilde{B}}=0$. The other two equations,
which contain $\psi_A$ and $\psi_{\tilde{A}}$, are written by
\begin{eqnarray}
\psi_{m+2A}+\beta(2 -\bar{\gamma_1}\bar{\gamma_3})\psi_{m+1A}
+(\beta^2-\bar{\gamma_1}\bar{\gamma_3})\psi_{mA}=0\\ \nonumber
\bar{\gamma_1}\psi_{m\tilde{A}}=-(\beta
\psi_{mA}+\psi_{m+1A})\qquad\qquad \nonumber \label{semiinfinite}
\end{eqnarray}
where $\bar{\gamma_1}\equiv\gamma_1/\gamma_0$,
$\bar{\gamma_3}\equiv\gamma_3/\gamma_0$, and $\beta\equiv 2\cos
(k/2)$. The first equation can be solved by substituting an ansatz
of $\psi_{mA}\propto\mu^m$ into Eq. (6). As expected from the double
degeneracy of band structures of bilayer graphite at the
particle-hole symmetry ($\ve =0$), one can obtain two independent
solutions $\mu_{\pm}$ as follows
\begin{eqnarray}
\mu_{\pm}&=&-\beta\Big(1-\frac{\bar{\gamma_1}\bar{\gamma_3}}{2}\Big)\\
&&\pm\sqrt{\bar{\gamma_1}\bar{\gamma_3}(1-\beta^2)+\beta^2\big(\frac{\bar{\gamma_1}\bar{\gamma_3}}{2}\big)^2},\nonumber
\end{eqnarray}
where the localization lengths are given by $\xi_{\pm}=|\ln
|\mu_{\pm}||^{-1}$.

We denote the two solutions as $\psi_{mA}^{\pm}$ for $\mu_{\pm}$
respectively. In order to have a physical normalizable solution,
one needs to impose the following condition for $\mu$: $|\mu |<
1$. This condition restricts the range of the $k$ values
supporting the zero energy edge states as follows
\begin{eqnarray}
{\rm For}\,\, \psi_{mA}^+&,&\,\,
2\cos^{-1}\frac{\sqrt{1+\bar{\gamma_1}\bar{\gamma_3}}}{2}\leq k
\leq\pi  \\ \nonumber {\rm For}\,\, \psi_{mA}^-& &\begin{cases}
2\cos^{-1}\frac{\sqrt{1+\bar{\gamma_1}\bar{\gamma_3}}}{2}\leq k
\leq\frac{2\pi}{3}\\
2\cos^{-1}\frac{1-\bar{\gamma_1}\bar{\gamma_3}}{2}\leq k \leq\pi
.\end{cases} \label{forbidden}
\end{eqnarray}

At $k=\pi$, the edge state has a form $\psi_{mA}=(\pm
\sqrt{\bar{\gamma_1}\bar{\gamma_3}})^m$ and $\psi_{m\tilde{A}}=-(\pm
\sqrt{\bar{\gamma_1}\bar{\gamma_3}})^{m+1}/\bar{\gamma_1}$. Hence
the localization length of the semi-infinite Z-BGNR is finite at the
zone boundary, while the semi-infinite monolayer GNR is totally
localized at one edge for $k=\pi$. It is interesting to notice that
the Z-BGNRs also support a completely localized mode at
$k^*=2\cos^{-1}\sqrt{\bar{\gamma_1}\bar{\gamma_3}}/2$ at which
$\mu=0$.

In Fig. 2(b), the schematic diagram of the edge states are plotted
with respect to $k$,
%These are
%illustrated in FIG.2 in the vicinity of $\varepsilon=0$
where the two edge modes are represented by two different colors.
The yellow(pale gray) edge mode $\psi_{mA}^+$ exists from the
point $T_1$ at
$k=2\cos^{-1}\sqrt{1+\bar{\gamma_1}\bar{\gamma_3}}/2$ to the zone
boundary as observed in the monolayer GNR as well.
%This edge region is a little larger than that of the
%GNR(monolayer).\cite{R2,R3}
Remarkably, we notice that the red(dark gray) edge mode
$\psi_{mA}^-$ has a forbidden region between the Dirac point $D$
at $k=2\pi/3$ and the point $T_2$ at
$k=2\cos^{-1}(1-\bar{\gamma_1}\bar{\gamma_3})/2$. The existence of
the forbidden region is the unique feature of the Z-BGNR. We draw
two symmetric warped bands within this region, since the bands
should be continuous and particle-hole symmetric. By comparing to
the graphene bilayer system, the point $D$ corresponds to the
Dirac point while the points $T_1$ and $T_2$ correspond to the
three legs' Fermi points $L_1$,$L_2$ and $L_3$ of the graphene
bilayer which are indicated in the inset of Fig. 2(a).\cite{R25}
This correspondence can be explicitly shown by comparing these
points. The previous study on the graphene bilayer has shown that
the three legs are displaced by
$2\bar{\gamma_1}\bar{\gamma_3}/\sqrt{3}$ from the Dirac point $D$
and are placed at the points with triangular symmetry. If one
projects these leg points onto the axis of zigzag chirality, the
two leg points are located at
$k=2\pi/3+2\bar{\gamma_1}\bar{\gamma_3}/\sqrt{3}$ and
$k=2\pi/3-\bar{\gamma_1}\bar{\gamma_3}/\sqrt{3}$. These values
previously obtained by the $\vec{k}\cdot\vec{p}$ approximation
perfectly agree with the two points $T_1$ and $T_2$ in Eq. (8) for
small $\bar{\gamma_1}$ and $\bar{\gamma_3}$. We have confirmed the
above feature of the zero energy edge states and the warped bands
by numerically solving Eq. (5) for relatively wide Z-BGNR with
widths $N=100$ and $200$ as shown in the inset of Fig. 2. In order
to see the above feature more clearly, we have taken about five
times larger values for $\bar{\gamma_1}$ and $\bar{\gamma_3}$ than
the experimental ones. Our numerical results agree well with that
of the semi-infinite Z-BGNR.

Recently, Ryu {\em et. al.} suggested a criterion for the
existence of the edge states based on the bulk energy dispersion,
which may be applicable to a certain class of edge
states\cite{R28}. They have studied the following particle-hole
symmetric Hamiltonian
\begin{eqnarray}
H=\sum_{\vec{k}}
\mathbf{c}^\dag_{\vec{k}}\mathbf{h}_{\vec{k}}\mathbf{c}_{\vec{k}}=\sum_{\vec{k}}
\mathbf{c}^\dag_{\vec{k}}\big(\vec{R}(\vec{k})\cdot\vec{\sigma}\big)\mathbf{c}_{\vec{k}}
\end{eqnarray}
where
$\mathbf{c}_{\vec{k}}^\dag=(c^\dag_{\vec{k}\uparrow},c_{\vec{k}\downarrow})$
and $\vec{\sigma}$ is a three component vector of the Pauli
matrices. This type of Hamiltonian includes the Bogoliubov-de
Gennes and the GNR Hamiltonian as well. Their criterion for this
Hamiltonian to have zero-energy edge sates is that the closed
trajectory of $\vec{R} (\vec{k})$ being confined in a 2D plane
encloses the gap-closing point. We want to understand our result
of Z-BGNR based on the above criterion. By analyzing the 2D
graphite bilayer system using exact diagonalization method, one
can obtain four energy bands, which we take to be $|\vec{R}
(\vec{k})|$.\cite{R28} We have analyzed the behavior of one of
$|\vec{R} (\vec{k})|$'s and noticed that it goes to zero three
times at $k_y=2\cos^{-1}\sqrt{1+\bar{\gamma_1}\bar{\gamma_3}}/2$,
$2\pi/3$, and $2\cos^{-1}(1-\bar{\gamma_1}\bar{\gamma_3})/2$
respectively. We presume that these three points indicate the
boundaries of the three distinct topological sectors of $|\vec{R}
(\vec{k})|$ where the closed loops within
$2\pi/3<k_y<2\cos^{-1}(1-\bar{\gamma_1}\bar{\gamma_3})/2$ exclude
the gap-closing point. Hence we argue that it corresponds to the
edge mode of $\psi_{mA}^-$. Strictly speaking, our bilayer
graphene system is out of their scope, since the bilayer
Hamiltonian is described by the $4\times 4$ traceless hermitian
matrix. Furthermore, since only the magnitude of $\vec{R}
(\vec{k})$ are available, we are not able to directly confirm our
conjecture yet.

We now calculate the full electronic structures of the Z-BGNRs
with finite ribbon width $N$ by numerically solving Eq. (5). We
obtain the band structures of two finite width Z-BGNRs with
$N=4,5$ as shown in Fig. 3. In order to study the edge states and
the low-lying energy excitations analytically, we diagonalize the
$4N\times 4N$ matrix of Eq. (5) with $\varepsilon=0$. By taking
the determinant of the matrix to be zero, that is,
$[(2\cos(k/2))^2-\bar{\gamma_1}\bar{\gamma_3}]^{2N}=0$, one can
see that the band touches $\varepsilon=0$ at $\pm k^*$ independent
of $N$. In order to calculate the dispersion relations around
$(k,\varepsilon)=(k^*,0)$, we expand the determinant with respect
to $(k^*,0)$ and obtain the following results
\begin{eqnarray}
\ve\approx\begin{cases}\pm\sqrt{\frac{\bar{\gamma_1}\bar{\gamma_3}}{\bar{\gamma_1}^2+\bar{\gamma_3}^2}}(2(k-k^*))^N\quad&(N=\mathrm{odd})\\
\quad\pm\sqrt{\frac{1}{2}}(2(k-k^*))^N&(N=\mathrm{even})
\end{cases}
\end{eqnarray}
where $\ve$ is written in units of $\gamma_0$. These peculiar
dispersions are demonstrated in our numerical result as shown in
the insets of Fig. 3. When the ribbon width $N$ goes to infinity,
the energy dispersion becomes flattened leading to the localized
edge states. The similar behavior has been observed for the
monolayer GNR with $k^*$ being replaced with $k=\pi$. Unlikely
from the monolayer GNR, a small gap opens at the zone boundary.
The magnitude of the gap can be calculated approximately by
considering the weak coupling between those edge modes localized
at the different edges. We define the following four states which
are constructed from the two semi-infinite Z-BGNRs localized at
the opposite edges for $k=\pi$
\begin{eqnarray}
|L\rangle&=&\psi_{1A}\big(|1A\rangle-\bar{\gamma_3}|2\tilde{A}\rangle+\bar{\gamma_1}\bar{\gamma_3}|3A\rangle+\cdot\cdot\cdot\big)\\
\nonumber
|\tilde{L}\rangle&=&\psi_{1\tilde{A}}\big(|1\tilde{A}\rangle-\bar{\gamma_1}|2A\rangle+\bar{\gamma_1}\bar{\gamma_3}|3\tilde{A}\rangle+\cdot\cdot\cdot\big)\\
\nonumber
|R\rangle&=&\psi_{NB}\big(|NB\rangle-\bar{\gamma_3}|N-1\tilde{B}\rangle\\&&\qquad\qquad\qquad+\bar{\gamma_1}\bar{\gamma_3}|N-2B\rangle+\cdot\cdot\cdot\big)\nonumber\\
\nonumber
|\tilde{R}\rangle&=&\psi_{N\tilde{B}}\big(|1A\rangle-\bar{\gamma_3}|N-1B\rangle\\&&\qquad\qquad\qquad+\bar{\gamma_1}\bar{\gamma_3}|N-2\tilde{B}\rangle+\cdot\cdot\cdot\big)\nonumber,
\end{eqnarray}
where $|L\rangle$ and $|\tilde{L}\rangle$ represent the localized
states at the left edge, while $|R\rangle$ and $|\tilde{R}\rangle$
are those at the right edge. Using these four orthogonal states as
the basis, the matrix elements of the tight binding Hamiltonian
can be calculated leading to the following results
\begin{eqnarray}
\ve=\begin{cases}\qquad\qquad\pm(\bar{\gamma_1}\bar{\gamma_3})^{\frac{N}{2}}\qquad\qquad\qquad(\mathrm{N=even})\\
\pm\bar{\gamma_1}(\bar{\gamma_1}\bar{\gamma_3})^{\frac{N-1}{2}},\quad\pm\bar{\gamma_3}(\bar{\gamma_1}\bar{\gamma_3})^{\frac{N-1}{2}}
\quad(\mathrm{N=odd}).\end{cases}
\end{eqnarray}
For the even width ribbons, the two split edge bands in the first
quadrant converge to a single value
$\ve=(\bar{\gamma_1}\bar{\gamma_3})^{N/2}$ at the zone boundary,
while they remain to be split for the odd ones. This is due to the
reduced symmetry of the odd width ribbons compared to the even
ones.

According to the previous experiments, the band parameter
$\gamma_4$ was assumed to be much smaller than $\gamma_1$ and
$\gamma_3$ by an order of magnitude.\cite{R25,R29,R30,R31,R32} The
typical values for the hopping amplitudes are taken to be
$\bar{\gamma_1}=0.12$, $\bar{\gamma_3}=0.1$ and
$\bar{\gamma_4}=0.014$.\cite{R27} Hence it is of practice to
neglect the contribution of $\gamma_4$. Here we want to
investigate the effect of $\gamma_4$ on the low-lying energy
dispersion. Within the $\vec{k}\cdot\vec{p}$ approximation, we
notice that not only the Fermi energy $\ve=0$ has a downward shift
by $\Delta\varepsilon\approx-2
{\bar\gamma}_1{\bar\gamma}_3^2{\bar\gamma}_4$, but also the
particle-hole symmetry is broken upon including $\gamma_4$.

Figure 4. (a) and (b) show the band structures for $N=4,5$
obtained by numerically solving Eq. (5). One can observe the
downward shift of the Fermi energy as expected. Interestingly, the
degenerate two bands at $k=k^*$ for $\gamma_4=0$ are split into
the two bands with an indirect energy gap. We denote the extremal
points as $k^e$ for electron pocket and $k^h$ for hole pocket as
indicated in the insets of Fig. 4. When $N$ goes to infinity, the
bands become flattened.

We have closely examined the positions of $k^e$ and $k^h$ with
increasing $N$ and observed that the inclusion of $\gamma_4$ leads
to a qualitatively different behavior. In Fig. 5, we plot $k^e$ and
$k^h$ with respect to $1/N$ and extrapolate to $N\rightarrow
\infty$. With increasing $N$, $k^e$ and $k^h$, which used to be
fixed at $k^*$, are moving towards the vicinity of the Dirac point
at $k=2\pi/3$. By fitting the energy dispersions near $k^*_e$ and
$k^*_h$ up to $N=4$, we find that the peculiar dispersions of
$\ve\sim (k-k^*)^N$ reduce to the parabolic one except for the $N=1$
case.

\section{Magnetic properties of the Z-BGNR}
When the Coulomb interaction $U$ is present in the monolayer
graphene system, sublattice ferromagnetism can appear beyond a
certain critical value of $U$. The most favorable spin
configuration is known to be the opposite magnetization between
two sublattices $A$ and $B$, which is consistent with Lieb's
theorem regarding the magnetism of bipartite lattices.\cite{R33}
For the case of the Z-GNR, the existence of almost flat bands at
zero-energy stabilizes the sublattice edge ferromagnetism even for
the weak Coulomb interaction. Recently it has been proposed that
the Z-GNR can become half-metallic when the transverse electric
field is applied.\cite{R17,R18,R22} We have studied the edge
magnetism for Z-BGNR by applying the Hartree-Fock approximation to
the half-filled Hubbard Hamiltonian. Following the Hartree-Fock
decoupling, the Hubbard Hamiltonian can be written by
\begin{eqnarray}
H_{MF}=-\sum_{\langle
i,j\rangle,\sigma}t_{ij}c^\dag_{i\sigma}c_{j\sigma}+U\sum_{i}\bigg[\langle
n_{i\downarrow}\rangle n_{i\uparrow}+\langle n_{i\uparrow}\rangle
n_{i\downarrow}-\langle n_{i\uparrow} \rangle \langle
n_{i\downarrow}\rangle\bigg]
\end{eqnarray}
where we have included the on-site Coulomb interaction $U$ and
$n_{i\sigma}=c^\dag_{i,\sigma}c_{i,\sigma}$. The mean electron
density $\langle n_{i\sigma}\rangle$ is kept to be constant in a given
dimer line due to the translational invariance.
%Although the
%sublattices ($A$, $\tilde{B}$) and ($B$, $\tilde{A}$) are not
%equivalent, the on-site Coulomb energy U is assumed to be same on
%every sites neglecting minor quantitative differences.
Based on the above Hartree-Fock Hamiltonian, we have solved the
self-consistent equations. Since the inter-layer hopping parameters
$\gamma_1$ and $\gamma_3$ are much smaller than the intra-layer one
$\gamma_0$ by an order of magnitude, we take the initial
configuration for iteration as the two coupled ground state
configurations of the monolayer GNR. Following the symmetry of
Z-BGNR, there exist two possible inter-layer spin configurations as
shown in Fig. 6. One is the anti-ferromagnetic alignment(AFM) of
inter-layer edge spins within the same side of edges: $\langle
n_{A,m,\sigma}\rangle =\langle n_{\tilde{B},N-m+1,\sigma}\rangle$
and $\langle n_{B,m,\sigma}\rangle =\langle
n_{\tilde{A},N-m+1,\sigma}\rangle$. The other is the ferromagnetic
alignment(FM) of inter-layer edge spins: $\langle
n_{A,m,\sigma}\rangle =\langle
n_{\tilde{B},N-m+1,{\bar\sigma}}\rangle$ and $\langle
n_{B,m,\sigma}\rangle =\langle
n_{\tilde{A},N-m+1,{\bar\sigma}}\rangle$.

In order to see which configuration is energetically favorable, we
have calculated the total energy of the system in those two cases.
The differences between two energies $E_{AFM}-E_{FM}$ are plotted as
a function of $U$ in Fig. 7. Here we have examined a Z-BGNR of width
$N=10$ and the inter-layer hopping parameters
$\bar{\gamma_1}=\bar{\gamma_3}=0.1$. We notice that both AFM and FM
configurations open a gap by lifting off the flat bands. Figure 7
demonstrates that the FM configuration becomes immediately stable
upon including the Coulomb interaction. Since the ferromagnetic
alignment of the upper and lower edge spins leads to the inter-layer
anti-ferromagnetic coupling between the nearest neighbor spin pairs
connected by $\gamma_1$ and $\gamma_3$, the ferromagnetic alignment
will become energetically favorable for large $U$. Hence for the
experimentally relevant value of $U$, that is, $U/\gamma_0\sim 1$,
the FM configurations are stable.\cite{R34,R35,R36} We have also
studied the bulk magnetism for the bilayer graphene system. In the
inset of figure 7, the energy differences $E_{AFM}-E_{FM}$ are
plotted as a function of $U$. With the increase of $U$ above $U_{c}
\cong 2.5\gamma_0$, the paramagnetic spin state becomes the FM one,
which agrees with the result of the monolayer graphene system. Hence
for the Z-BGNR, the spin fluctuations at the edges enhance the
appearance of the FM configurations even below $U_{c}$.

\section{Conclusions}
We have studied the electronic and magnetic structures of edge
states of the zigzag bilayer graphite nanoribbons within the tight
binding approximation. Neglecting the band parameter $\gamma_4$, we
obtained a peculiar dispersion relation $\ve\sim\pm(k-k^*)^N$ with
$k^*=2\cos^{-1}\sqrt{\bar{\gamma_1}\bar{\gamma_3}}/2$ which is fixed
independent of the ribbon width $N$. As $N$ increases, the
dispersionless edge bands at $\ve=0$ will appear just like a
monolayer zigzag GNR. By investigating the semi-infinite Z-BGNR
analytically, we notice that the trigonal warping of the bilayer
graphene leads to the following interesting effect on the edge state
structure of the Z-BGNR. One of the edge state at $\ve=0$ is absent
within the region between Dirac point($k=2\pi/3$) to
$k=2\cos^{-1}(1-\bar{\gamma_1}\bar{\gamma_3}/2)$, where the bulk
warped bands used to be present. With the inclusion of $\gamma_4$,
the particle-hole symmetry is broken in our Z-BGNR system. There
exists an indirect gap between two edge bands, whose band extrema
are located at $k^e$ and $k^h$ respectively. Instead of the fixed
$k^*$ as for the case of $\gamma_4=0$, $k^e$ and $k^h$ approaches
near to the Dirac point with increasing $N$. The magnetic property
of the Z-BGNR is studied by solving the Hubbard type Hamiltonian
using Hartree-Fock approximation. We have shown that for the
realistic value of $U$, the ferromagnetic alignment of inter-layer
edge spins within the same side of edges is favored.

\begin{acknowledgments}
We thank Y. Son and J. Yu for valuable discussions. This work was
supported by the Korea Research Foundation Grant funded by the
Korean Government (MOEHRD, Basic Research Promotion Fund) through
KRF-2006-311-C00286. J.R. acknowledges partial support from Seoul
city through the program Seoul Science Scholarship.
\end{acknowledgments}

\clearpage
\newpage
\begin{center}
\begin{figure}
\caption{(Color online)(a) Our model system of Bernal stacked
Z-BGNR. Both the three inter-layer($\gamma_1$, $\gamma_3$ and
$\gamma_4$) and one intra-layer hopping($\gamma_0$) parameters are
shown. The Z-BGNR has a translational symmetry along $\vec{T}$.
The sublattice A is colored yellow(pale gray) while B is colored
green(dark gray). (b) The definition of the width of the Z-BGNR.
Here the red lines(pale gray) are the lower graphene sheet and the
black lines the upper one. The tilde above A and B stand for the
upper layer and the black box represents a unit cell of the
Z-BGNR.}
\end{figure}

\begin{figure}
\caption{(Color online)(a) Projected band structure of 2D graphite
bilayer along the direction of zigzag axis. The boxed region near
the Dirac point is magnified in the inset, which exhibits the
trigonal warping. $D$ is the Dirac point at $k=2\pi /3$ and three
$L$ points are the Fermi points of three nearby pockets. (b) A
schematic diagram of the zero-energy edge states of the
semi-infinite Z-BGNR near the Dirac point within the red box in Fig.
2(a). At $\ve=0$, the two eigenstates are drawn with different
colors. Red(dark gray) for $\psi_{-}$ and yellow(pale gray) for
$\psi_{+}$. $D$ is the Dirac point while $T_1$, $T_2$, and the two
warped bands reflect on the effect of the trigonal warping of the
graphite bilayer. The inset shows the energy dispersion curve
obtained by numerical method for the Z-BGNRs with finite widths
$N=100,200$.}
\end{figure}

\begin{figure}
\caption{(Color online)(a)The band diagram of the Z-BGNR with the
following parameters $\bar{\gamma_1}=0.12$, $\bar{\gamma_3}=0.1$
and $\bar{\gamma_4}=0$ for $N=4$. The inset, a magnified view of
the low energy region, shows the power law dispersion of the edge
bands. (b) The band diagram of the Z-BGNR for $N=5$.}
\end{figure}

\begin{figure}
\caption{(Color online)(a)The band diagram for the Z-BGNR with the
band parameters $\bar{\gamma_1}=0.12$, $\bar{\gamma_3}=0.1$ and
$\bar{\gamma_4}=0.014$ for $N=4$. The inset, a magnified view of
the low energy region, demonstrates the broken particle-hole
symmetry and the parabolic dispersions. (b) The band diagram of
the Z-BGNR for $N=5$.}
\end{figure}

\begin{figure}
\caption{(Color online) The plot of $k^e$ and $k^h$ with respect
to $1/N$. The extrapolated values of $k^e$ and $k^h$ approach near
to the Dirac point ($k=2\pi/3\sim 2.094$).}
\end{figure}

\begin{figure}
\caption{(Color online) Two possible configurations of the
magnetic moments are illustrated at the left part of the Z-BGNR.
Black and gray lines represent dimer lines of the upper and lower
layer each. Yellow(pale gray) arrows are the magnetic moments of
the lower layer while red(dark gray) arrows for upper one.  (a)
The AFM configuration. At the left edge, the spin polarization at
upper edge($1\tilde{A}$) and lower edge($1A$) layer are opposite.
(b) The FM configuration. The spin polarization at upper
edge($1\tilde{A}$) and lower edge($1A$) layer are parallel.}
\end{figure}

\begin{figure}
\caption{(Color online) A plot of the energy differences per unit
cell between the AFM and FM configurations for the N=10 Z-BNGR and
2D graphite bilayer(inset) plotted as a function of U. The unit of
the energy is $\gamma_0$.}
\end{figure}
\end{center}

%\newpage
%\begin{table}
%\caption{Caption Caption  Caption  Caption.}
%\begin{ruledtabular}
%\begin{tabular}{cccc}
% 1 & 2 & 3 & 4 \\
% 2 & 4 & 6 & 8
%\end{tabular}
%\end{ruledtabular}
%\end{table}

%\newpage
%\begin{figure}[t!]
%\includegraphics[width=13.0cm]{submitsamplef1.png}
%\caption{caption Caption Caption Caption.} \label{fig.1}
%\end{figure}%

%\newpage
%\begin{figure}[t!]
%\includegraphics[width=13.0cm]{submitsamplef2.png}
%\caption{caption Caption Caption Caption.} \label{fig.2}
%\end{figure}


\begin{references}
%\begin{thebibliography}{}
\bibitem{R1} Melinda Y. Han, Barbaros Ozyilmaz, Yuanbo Zhang, and Philip Kim, Phys. Rev. Lett. \textbf{98}, 206805 (2007).
\bibitem{R2} Li Yang, Cheol-Hwan Park, Young-Woo Son, Marvin L. Cohen and Steven G. Louie, Phys. Rev. Lett. \textbf{99}, 186801 (2007).
\bibitem{R3} Mitsutaka Fujita, Katsunori Wakabayashi, Kyoko Nakada and Koichi Kusakabe, J. Phys. Soc. Jpn. {\bf 65}, 1920 (1996).
\bibitem{R4} F. L. Shyu, Ming Fa Lin, C. P. Chang, R. B. Chen, J. S. Shyu, Y. C. Wang and c. H. Liao, J. Phys. Soc. Jpn. {\bf 70}, 3348 (2001).
\bibitem{R5} Yoshiyuki Miyamoto, Kyoko Nakada, and Mitsutaka Fujita, Phys. Rev. B \textbf{59}, 9858 (1999).
\bibitem{R6} Takazumi Kawai, Yoshiyuki Miyamoto, Osamu Sugino, and Yoshinori Koga, Phys. Rev. B \textbf{62}, 16349(R) (2000).
\bibitem{R7} L. G. Cancado, M. A. Pimenta, B. R. A. Neves, G. Medeiros-Ribeiro, Toshiaki Enoki, Yousuke Kobayashi, Kazuyuki Takai, Ken-ichi Fukui, M. S. Dresselhaus, R. Saito, and A. Jorio, Phys. Rev. Lett. \textbf{93}, 047403 (2004).
\bibitem{R8} Y. Niimi, T. Matsui, H. Kambara, K. Tagami, M. Tsukada, and Hiroshi Fukuyama, Phys. Rev. B \textbf{73}, 085421 (2006).
\bibitem{R9} L. Brey and H. A. Fertig, Phys. Rev. B \textbf{73}, 235411 (2006).
\bibitem{R10} L. Brey and H. A. Fertig, Phys. Rev. B \textbf{75}, 125434 (2007).
\bibitem{R11} Huaixiu Zheng, Z. F. Wang, Tao Luo, Q. W. Shi, and Jie Chen, Phys. Rev. B \textbf{75}, 165414 (2007).
\bibitem{R12} Daniel Finkenstadt, G. Pennington and M. J. Mehl, Phys. Rev. B \textbf{76}, 121405(R) (2007).
\bibitem{R13} Mahito Kohmoto and Yasumasa Hasegawa, Phys. Rev. B \textbf{76}, 205402 (2007).
\bibitem{R14} Kyoko Nakada, Mitsutaka Fujita, Gene Dresselhaus, and Mildred S. Dresselhaus, Phys. Rev. B \textbf{54}, 17954 (1996).
\bibitem{R15} Katsunori Wakabayashi, Mitsutaka Fujita, Hiroshi Ajiki, and Manfred Sigrist, Phys. Rev. B \textbf{59}, 8271 (1999).
\bibitem{R16} Young-Woo Son, Marvin L. Cohen, and Steven G. Louie, Phys. Rev. Lett. \textbf{97}, 216803 (2006).
\bibitem{R17} L. Pisani, J. A. Chan, B. Montanari, and N. M. Harrison, Phys. Rev. B \textbf{75}, 064418 (2007).

\bibitem{R18} Young-Woo Son, Marvin L. Cohen, and Steven G. Louie, Nature \textbf{444}, 347 (2006).
\bibitem{R19} D. J. Klein, Chem. Phys. Lett. \textbf{217}, 261 (1994).

\bibitem{R20} J. Fernandez-Rossier and J. J. Palacios, Phys. Rev. Lett. \textbf{99}, 177204 (2007).
\bibitem{R21} Oded Hod, Juan E. Peralta, and Gustavo E. Scuseria, Phys. Rev. B \textbf{76}, 233401 (2007).
\bibitem{R22} Koichi Kusakabe and Masanori Maruyama, Phys. Rev. B \textbf{67}, 092406 (2003).
\bibitem{R24} Hosik Lee, Noejung Park, Young-Woo Son, Seungwu Han, and Jaejun Yu, Chem. Phys. Lett. \textbf{398}, 207
(2004); Hosik Lee, Young-Woo Son, Noejung Park, Seungwu Han, and
Jaejun Yu, Phys. Rev. B \textbf{72}, 174431 (2005).

\bibitem{R25} Edward McCann and Vladimir I. Fal'ko, Phys. Rev. Lett. \textbf{96}, 086805 (2006).
\bibitem{R26} J$\acute{\mathrm{o}}$zsef Cserti, Andr$\acute{\mathrm{a}}$s Csord$\acute{\mathrm{a}}$s, and Gyula D$\acute{\mathrm{a}}$vid, Phys. Rev. Lett. \textbf{99}, 066802 (2007).
\bibitem{R27} J.-C. Charlier, X. Gonze, and J.-P. Michenaud, Phys. Rev. B \textbf{43}, 4579 (1991).
\bibitem{R28} Shinsei Ryu and Yasuhiro Hatsugai, Phys. Rev. Lett. \textbf{89}, 077002 (2002).
\bibitem{R29} Taisuke Ohta, Aaron Bostwick, Thomas Seyller, Karsten Horn, and Eli Rotenberg, Science {\bf 313}, 951 (2006).
\bibitem{R30} Taisuke Ohta, Aaron Bostwick, J. L. McChesney, Thomas Seyller, Karsten Horn, and Eli Rotenberg, Phys. Rev. Lett. \textbf{98}, 206802 (2007).
\bibitem{R31} Xue-Feng Wang and Tapash Chakraborty, Phys. Rev. B \textbf{75}, 041404 (2007).
\bibitem{R32} E. Mendez, A. Misu, and M. S. Dresselhaus, Phys. Rev. B \textbf{21}, 827 (1980).

\bibitem{R33} E. H. Lieb, Phys. Rev. Lett. \textbf{62}, 1201 (1989).
\bibitem{R34} U. von Barth and L. Hedin, J. Phys. C \textbf{5}, 1629 (1972).
\bibitem{R35} J. P. Perdew, K. Burke and M. Ernzerhof, Phys. Rev. Lett. \textbf{77}, 3865 (1996).
\bibitem{R36} E. V. Castro, N. Peres, J. Santos, Phys. Rev. Lett. \textbf{100}, 026802 (2008).


%\end{thebibliography}
\end{references}
\end{document}